\documentclass[aps,prc,twocolumn,superscriptaddress,showpacs,
nofootinbib]{revtex4-1}
\usepackage{hyperref}
\usepackage{bm}
\usepackage[all]{hypcap}
\usepackage{graphicx,amsmath,amssymb,bm,tabularx,dsfont}
\usepackage[usenames,dvipsnames,svgnames,table]{xcolor}
\usepackage{enumerate}
\usepackage{isotope}
\usepackage{multirow,dcolumn}
\usepackage{bbm}
\usepackage{braket}
\usepackage{textcomp}
\usepackage{gensymb}

\newcolumntype{.}{D{.}{.}{2.3}}
\newcolumntype{,}{D{.}{.}{1.1}}

\hypersetup{colorlinks=true,citecolor=[rgb]{0,0,0.8},
linkcolor=[rgb]{0,0,0.8},urlcolor=[rgb]{0,0,0.8}}

\allowdisplaybreaks[1]

\newcommand{\mev}{~\text{MeV}}

\newcommand{\fm}{~\mathrm{fm}}

\newcommand{\beq}{\begin{equation}}
\newcommand{\eeq}{\end{equation}}

\newcommand{\fet}[1]{\mbox{\boldmath $#1$}}
\definecolor{dartmouth}{rgb}{0.05,0.5,0.06}

\usepackage[normalem]{ulem}
\definecolor{joelcolor}{rgb}{0.90,0.40,0.40}

\allowdisplaybreaks

\begin{document}

\title{Large-cutoff behavior of local chiral effective field theory interactions}
\author{I.\ Tews}
\email[E-mail:~]{itews@uw.edu}
\affiliation{Institute for Nuclear Theory,
University of Washington, Seattle, Washington, 98195-1550, USA}
\affiliation{JINA-CEE, Michigan State University, 
East Lansing, Michigan, 48823, USA}
\author{L.\ Huth}
\email[E-mail:~]{lukashuth@theorie.ikp.physik.tu-darmstadt.de}
\affiliation{Institut f\"ur Kernphysik, Technische Universit\"at Darmstadt, 64289 Darmstadt, Germany}
\affiliation{ExtreMe Matter Institute EMMI, GSI Helmholtzzentrum f\"ur
Schwerionenforschung GmbH, 64291 Darmstadt, Germany}
\author{A.\ Schwenk}
\email[E-mail:~]{schwenk@physik.tu-darmstadt.de}
\affiliation{Institut f\"ur Kernphysik, Technische Universit\"at Darmstadt, 64289 Darmstadt, Germany}
\affiliation{ExtreMe Matter Institute EMMI, GSI Helmholtzzentrum f\"ur
Schwerionenforschung GmbH, 64291 Darmstadt, Germany}
\affiliation{Max-Planck-Institut f\"ur Kernphysik, Saupfercheckweg 1, 
69117 Heidelberg, Germany}

\begin{abstract}
Interactions from chiral effective field theory have been successfully 
employed in a broad range of \textit{ab initio} calculations of nuclei and nuclear 
matter, but it has been observed that most results of few- and many-body calculations experience a substantial residual regulator and cutoff dependence. 
In this work, we investigate the behavior 
of local chiral potentials at leading order under variation of 
the cutoff scale for different local regulators. 
When varying the cutoff, we require that the resulting interaction produces no spurious bound states in the deuteron channel. 
We find that, for a particular choice of leading-order operators, 
nucleon-nucleon phase shifts and the deuteron ground-state energy 
converge to cutoff-independent plateaus for all regulator functions
we investigate. This observation may enable improved calculations 
with chiral Hamiltonians that also include three-nucleon interactions.
\end{abstract}

\maketitle

\section{Introduction}

Chiral effective field theory (EFT)~\cite{Epelbaum:2008ga, Machleidt:2011zz} 
has been shown to be a powerful framework to derive nuclear interactions. It 
provides a systematic expansion for nuclear forces that is linked to the 
symmetries of quantum chromodynamics (QCD). This systematic expansion 
is based on a so-called power-counting (PC) scheme that ideally allows one 
to arrange different contributions to the interaction according to their 
importance. Thus, the PC scheme establishes a truncation scheme, which enables one to obtain systematic uncertainty estimates. 

Modern chiral EFT interactions are typically constructed by applying Weinberg 
power counting (WPC)~\cite{Weinberg1979,Weinberg1990, Weinberg1991, 
Weinberg1992}, which is based on dimensional analysis in momentum space. 
In the purely pionic and one-nucleon sector, due to the Goldstone-boson nature 
of pions, all amplitudes can be expanded in powers of the dimensionless 
expansion parameter $Q/\Lambda_b$, where $Q$ is a typical momentum 
of the system and $\Lambda_b$ is the breakdown scale of the theory. 
In the two-nucleon sector, instead, bound states appear and the problem 
becomes nonperturbative. To obtain observables, WPC suggests defining the 
nuclear potential as the sum of all irreducible diagrams that do not contain 
purely nucleonic intermediate states and thus are not infrared enhanced. This sum is then 
truncated according to a power counting in $Q/\Lambda_b$. The resulting 
potential is iterated to all orders by solving the Lippmann-Schwinger (LS) or 
Schroedinger equation. At the two-body level, at leading order (LO), WPC 
leads to the appearance of $S$-wave contact interactions and the one-pion-exchange (OPE) interaction, while at higher orders additional derivative contact interactions 
and multipion-exchange interactions, as well as corrections to previous topologies, have to be considered. 

This approach has unsolved problems. In few- and many-body calculations,
a regularization scheme has to be introduced to cut off high-momentum 
modes that may lead to divergences. Because the regularization scheme 
is arbitrary, results should be independent of this choice after the dependence 
of contact parameters on the regularization scale, the so-called cutoff 
$\Lambda_c$, is taken into account. This means that at each order there
should be sufficiently many counterterms to absorb any residual cutoff 
dependence in the limit $\Lambda_c \to \infty$. This is problematic for 
singular potentials, such as OPE, which has a $1/r^3$ behavior in spin $S=1$ 
channels due to the tensor force. Although the focus is to describe 
long-range behavior, this singular potential nevertheless generates an 
oscillatory wave 
function for $r\to 0$ that leads to the appearance of spurious bound states 
and cutoff-dependent results. To renormalize such a potential in a certain 
partial wave, i.e., to obtain cutoff-independent results for large cutoffs, a counterterm is 
necessary in the same partial wave~\cite{Beane:2000wh, Nogga:2005hy}. 
In WPC at LO, however, the only counterterms appear in the $S$ waves but 
not in partial waves with orbital angular momentum $l>0$, where the singular OPE potential also contributes.

Kaplan, Savage, and Wise (KSW)~\cite{Kaplan:1998tg, Kaplan:1998we}  
suggested a different PC that uses dimensional regularization 
(DR) with power divergence subtraction to systematically expand the 
nucleon-nucleon ($NN$) scattering amplitude in powers of $Q/\Lambda_b$ 
instead of the potential. Within this scheme, only the LO contact 
interactions are treated nonperturbatively, while other contact interactions and 
pion exchanges are treated in finite order in perturbation theory, in order to 
find analytic expressions for the scattering amplitudes. Although the KSW 
power counting is well defined and consistent, it failed to reproduce the 
phase shifts from the Nijmegen partial-wave analysis (PWA)~\cite{Stoks:1993tb}  
in spin-triplet channels at next-to-next-to-leading order (N$^2$LO)~\cite{Fleming:1999ee} and led to large 
N$^2$LO corrections. It was found that in some 
spin-triplet channels, the nonperturbative treatment of pion-exchange 
diagrams is necessary at higher momenta because 
the OPE tensor force is large and singular and must be summed to 
all orders. This is done in WPC and reflects the fact that a correct 
renormalization of singular potentials is intrinsically
nonperturbative~\cite{Beane:2001bc}. A solution to this problem was
suggested in Ref.~\cite{Beane:2008bt}, where the short-range part
of the OPE interaction was canceled by fictitious heavy mesons. 

Nogga, Timmermans, and van Kolck (NTvK)~\cite{Nogga:2005hy} studied the 
cutoff dependence of phase shifts at LO in WPC for nonlocal regulators and for 
a cutoff range of $\Lambda_c=2-20 \fm^{-1}$. They found that WPC leads to 
cutoff-independent results in the spin $S=0$ channels, in the $^3S_1$ channel, 
where WPC includes a counterterm, and in $S=1$ channels with repulsive 
tensor forces, e.g., $^3P_1$. However, NTvK observed strong cutoff
dependences and the appearance of several spurious bound states in the other 
attractive tensor channels, $^3P_0, ^3D_2$, and $^3P_2-^3F_2$, where 
there are no counterterms present in WPC. As a solution, NTvK suggested to 
explicitly add counterterms to partial waves with attractive tensor interactions,
i.e., $^3P_0$, $^3P_2$, and $^3D_2$ (see also, e.g., Refs.~\cite{PavonValderrama:2005wv,%
PavonValderrama:2005uj}). In higher partial waves, NTvK found 
that the centrifugal barrier screened the singular tensor force sufficiently
so that no counterterms were necessary in the investigated cutoff range. 
The modification of WPC proposed by NTvK was confirmed based on a
renormalization group analysis~\cite{Birse:2005um}, but also triggered
a debate on whether it remains necessary in higher order descriptions of $NN$
scattering; see, e.g., Ref.~\cite{Epelbaum:2006pt}.

In this paper, we investigate the large-cutoff 
behavior of local chiral interactions that were introduced in 
Refs.~\cite{Gezerlis:2013ipa, Gezerlis:2014zia}. It has been found that 
local regulators induce regulator artifacts, which mix contact interactions 
in a certain partial wave into all 
higher partial waves~\cite{Lynn:2015jua, Dyhdalo:2016ygz, Huth:2017wzw}, 
because the 
regulator does not commute with the antisymmetrizer. These regulator artifacts 
have been analyzed in detail in Ref.~\cite{Huth:2017wzw} for chiral $NN$ 
interactions at LO. In this work, we exploit these regulator artifacts to mix LO 
contact interaction terms into all attractive tensor channels to obtain 
cutoff-independent results for the phase shifts and deuteron ground-state
energy. 

This paper is organized as follows. In Sec.~\ref{sec:LOinteractions}, we 
introduce the local chiral interactions at LO that we consider in this work. 
In Sec.~\ref{sec:phaseshifts}, we then present the results for the phase 
shifts and the deuteron ground-state energy for these interactions and discuss 
the results in Sec.~\ref{sec:discussion}. Finally, we summarize in Sec.~\ref{sec:summary}.

\section{Local chiral interactions at LO}\label{sec:LOinteractions}

At LO, local chiral potentials in coordinate space are given by
\begin{align}
V_{\text{NN}}^{\text{LO}}(r,R_0)=V_{\text{OPE}}^{\text{LO}}(r,R_0)+V^{\text{LO}}_{\text{cont}}(r,R_0)\,,
\end{align}
with the OPE interaction
\begin{align}
& V_{\text{OPE}}^{\text{LO}}(r, R_0)= \frac{M_{\pi}^3}{12 \pi} \left(\frac{g_A}{2
F_{\pi}}\right)^2 \,\frac{e^{-M_{\pi} r}}{M_{\pi} r}\\
&\quad \times \left(  \sigma_{12}  + \left(1+\frac{3}{M_{\pi} r}+\frac{3}{(M_{\pi} r)^2} \right) \; S_{12}\right) \, \tau_{12} \, f_{\text{l}}(r, R_0)\,,\nonumber
\end{align}
and the general set of contact interactions at LO,
given by
\begin{align}
\label{eq:Loc_Cont}
V^{\text{LO}}_{\text{cont}}(r,R_0) &= \left( C_{\mathrm{1}} + C_{\sigma} \,\sigma_{12} +C_{\tau} \,\tau_{12} \right. \\
\nonumber &\quad \left. +C_{\sigma \tau} \,\sigma_{12} \,\tau_{12} \right ) f_{\text{s}}(r, R_0)\,,
\end{align}
where $\sigma_{12}=\fet \sigma_1\cdot \fet \sigma_2$ and similar for $\tau_{12}$.
The contact interactions are generally fit to $S$-wave $NN$ 
scattering. We use the long-range and short-range local regulator 
functions $f_l$ and $f_s$,
\begin{align}
f_{\text{l}}(r, R_0)&= \left(1-\exp\left(-\left(\frac{r}{R_0}\right)^{n_1} \right)\right)^{n_2}\,,\\
f_{\text{s}}(r, R_0)&= \frac{n}{4 \pi \Gamma(3/n) R_0^3}\exp\left(-\left(\frac{r}{R_0}\right)^n \right)\,,
\label{eq:regulator}
\end{align}
with the coordinate-space cutoff $R_0$, and where $n_1$, $n_2$, and $n$
determine the width of the regulator functions. In this work, we will investigate different combinations of 
$n_1$, $n_2$, and $n$ for these regulator functions, as low-energy 
physics should
be independent of the short-range details and any regulator function should 
be equally valid~\cite{Beane:2000wh}.

\begin{figure*}[t]
\includegraphics[trim= 0.1cm 0 0 0, clip=,width=1.\textwidth]{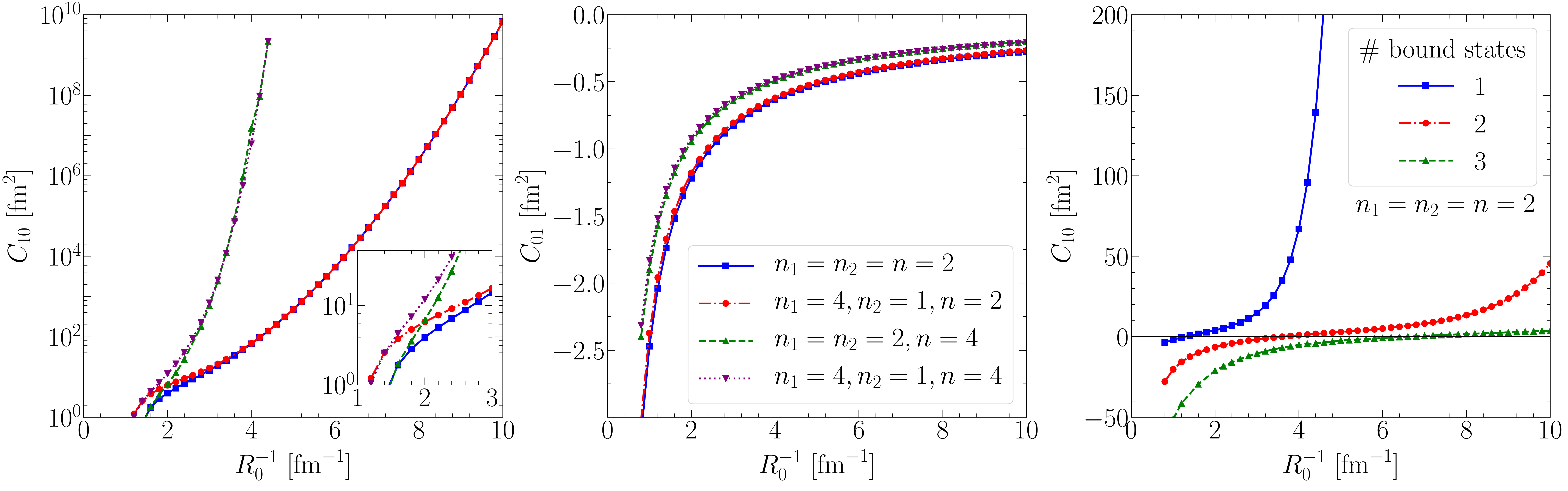}
\caption{\label{fig:para}
Spin-isospin LECs $C_{10}$ (left panel, logarithmic scale) and $C_{01}$ 
(middle panel) as functions of the inverse cutoff $R_0^{-1}$ for local chiral 
interactions at LO with different local regulators characterized by $n_1$,
$n_2$, $n$. Right panel: Spin-isospin LEC $C_{10}$ as function of the inverse 
cutoff $R_0^{-1}$ for local chiral interactions at LO allowing one, two, 
and three bound states in the coupled $^3S_1$--$^3D_1$ channel.
}
\end{figure*} 

When the regulator commutes with the antisymmetrizer, as is the case for 
typical nonlocal regulators, only two of the four operators in 
Eq.~\eqref{eq:Loc_Cont} are linearly independent, and one can choose any
two of the four operator structures  for the LO potential 
(except $\mathds{1},\sigma_{12}\tau_{12}$ 
which is linearly dependent in the two $S$-wave channels). 
This is known as the Fierz ambiguity. From the spin-isospin LECs $C_{ST}$, 
which enter partial waves with spin $S$ and the isospin $T$, one 
can then determine the operator LECs for different operator pairs according 
to
\begin{align}
\begin{pmatrix} C_{00}\\ C_{01} \\ C_{10} \\ C_{11} \end{pmatrix} 
= \begin{pmatrix} 1 & -3 & -3 & \hphantom{-}9 \\ 1 & -3 & \hphantom{-}1 & -3 \\ 1 & \hphantom{-}1 & -3 & -3 \\ 1 & \hphantom{-}1 & \hphantom{-}1 & \hphantom{-}1 \end{pmatrix}
\begin{pmatrix} C_{\mathbbm{1}\hphantom{\sigma}}\\ C_{\sigma\hphantom{\sigma}} \\ C_{\tau\hphantom{\sigma}} \\ C_{\sigma \tau} \end{pmatrix}\,.
\label{eq:LECtrafo}
\end{align}
However, the Fierz ambiguity is violated when local regulators are chosen; 
see Ref.~\cite{Huth:2017wzw} for a detailed discussion. 
In this case, when choosing two out of the 
four operators, the regulator affects the partial-wave decomposition and  
regulator artifacts appear in all higher partial waves. These artifacts have 
the form of higher-order contact operators and their LECs 
depend on the LO operator choice. While this mixing of LO $S$-wave 
physics into higher partial 
waves cannot be turned off completely, one can construct interactions that 
vanish in the $ST = 00$ and $ST =11$ partial waves, by choosing all 
four operators at LO and requiring $C_{00}=C_{11}=0$. 
We denote the latter interaction LO$_{{\rm n}P}$, and it is closest to 
nonlocally regularized interactions.
 
In this work, we construct various potentials for different operator structures, 
different sets $\lbrace n_1,n_2,n\rbrace$, and different cutoff values in 
the range $R_0^{-1} = 0.8-10.0\ \mathrm{fm}^{-1}$ 
in steps of $0.2\ \mathrm{fm}^{-1}$, where we use the inverse of the 
coordinate-space cutoff because it is connected to the momentum-space 
cutoff. Then, large values for $R_0^{-1}$ imply large momentum 
cutoffs. For each potential, we fit the two spin-isospin LECs in the $^1S_0$ 
and $^3S_1$ partial waves, $C_{01}$ and $C_{10}$, to the corresponding 
$S$-wave phase shifts from the Nijmegen partial-wave analysis 
(PWA)~\cite{Stoks:1993tb} up to laboratory energies of $50\mev$, as in 
Ref.~\cite{Huth:2017wzw}. When varying $R_0$, we renormalize 
the LECs $C_{ST}=C_{ST}(R_0)$ to keep the phase shifts invariant at 
low energies. Depending on the size and sign of the LECs, we can obtain 
fits with an arbitrary number of bound states. Therefore, when fitting the 
LECs, we additionally require that the resulting interaction allows for exactly 
one bound state in the deuteron channel. 

We present the resulting LECs $C_{10}$ 
and $C_{01}$ as functions of the inverse cutoff $R_0^{-1}$ in the first two 
panels of Fig.~\ref{fig:para}, respectively, for four different sets 
$\lbrace n_1,n_2,n\rbrace$ = $\lbrace 2,2,2\rbrace$, 
$\lbrace 4,1,2\rbrace$, $\lbrace 2,2,4\rbrace$, and $\lbrace 4,1,4\rbrace$. 
We find that the LEC dependence on the cutoff is 
regulator dependent, see also Ref.~\cite{Nogga:2005hy}, and observe a 
systematic behavior for the different regulator choices in both $ST$ channels.

We observe a strong increase of $C_{10}$ with $R_0^{-1}$ (in the left 
panel of Fig.~\ref{fig:para}), which is 
more prominent for short-range regulators with $n=4$ (green and purple lines) 
compared to those with $n=2$ (blue and red lines). This increase is due to the 
increasing attractive OPE tensor contribution in the $ST=10$ channel 
when increasing $R_0^{-1}$. Because we only allow for one 
bound state in the deuteron channel, an increasing LEC is needed to balance 
the OPE attraction sufficiently so that no second bound state enters. We will discuss this in more detail later in this paper. 
For the interactions with $n=4$, this requires a larger LEC 
because the regulator is sharper. Also, for $n=4$ we could not achieve a fit
beyond $R_0^{-1} = 4.4\ \mathrm{fm}^{-1}$, due to large numerical values
and cancellations. 

In the $ST=10$ channel, at small values of $R_0^{-1}$, we find pairwise 
similar LECs for the sets $\lbrace 2,2,2\rbrace$ (blue line) and 
$\lbrace 2,2,4\rbrace$ (green line) as well as for the sets 
$\lbrace 4,1,2\rbrace$ (red line) and 
$\lbrace 4,1,4\rbrace$ (purple line). At large values of $R_0^{-1}$ this 
behavior changes, and we find pairwise similar LECs for the sets 
$\lbrace 2,2,2\rbrace$ (blue line) and 
$\lbrace 4,1,2\rbrace$ (red line), and for the sets 
$\lbrace 2,2,4\rbrace$ (green line) and $\lbrace 4,1,4\rbrace$ (purple line). 
From this, we can deduce that at small $R_0^{-1}$, the LEC $C_{10}$ is 
dominated by effects of the long-range regulator, while at large $R_0^{-1}$, 
the LEC is dominated by effects of the short-range regulator. The transition 
between those two regimes is found at cutoffs $R_0^{-1}\approx 2 \fm^{-1}$. 
This transition region is shown in the inset in the left panel of 
Fig.~\ref{fig:para}. At small $R_0^{-1}$, the OPE is cut off at larger 
distances, 
and the LECs are rather small, so that the relative importance of the OPE is 
larger and the fits are mostly sensitive to the form of the long-range 
regulator.  This changes at large $R_0^{-1}$, where differences in the 
long-range regulator are not so important because the short-range interactions 
cut off the OPE at a certain distance scale, independent of the long-range 
regulator function.

\begin{figure*}[t]
\includegraphics[width=1\textwidth]{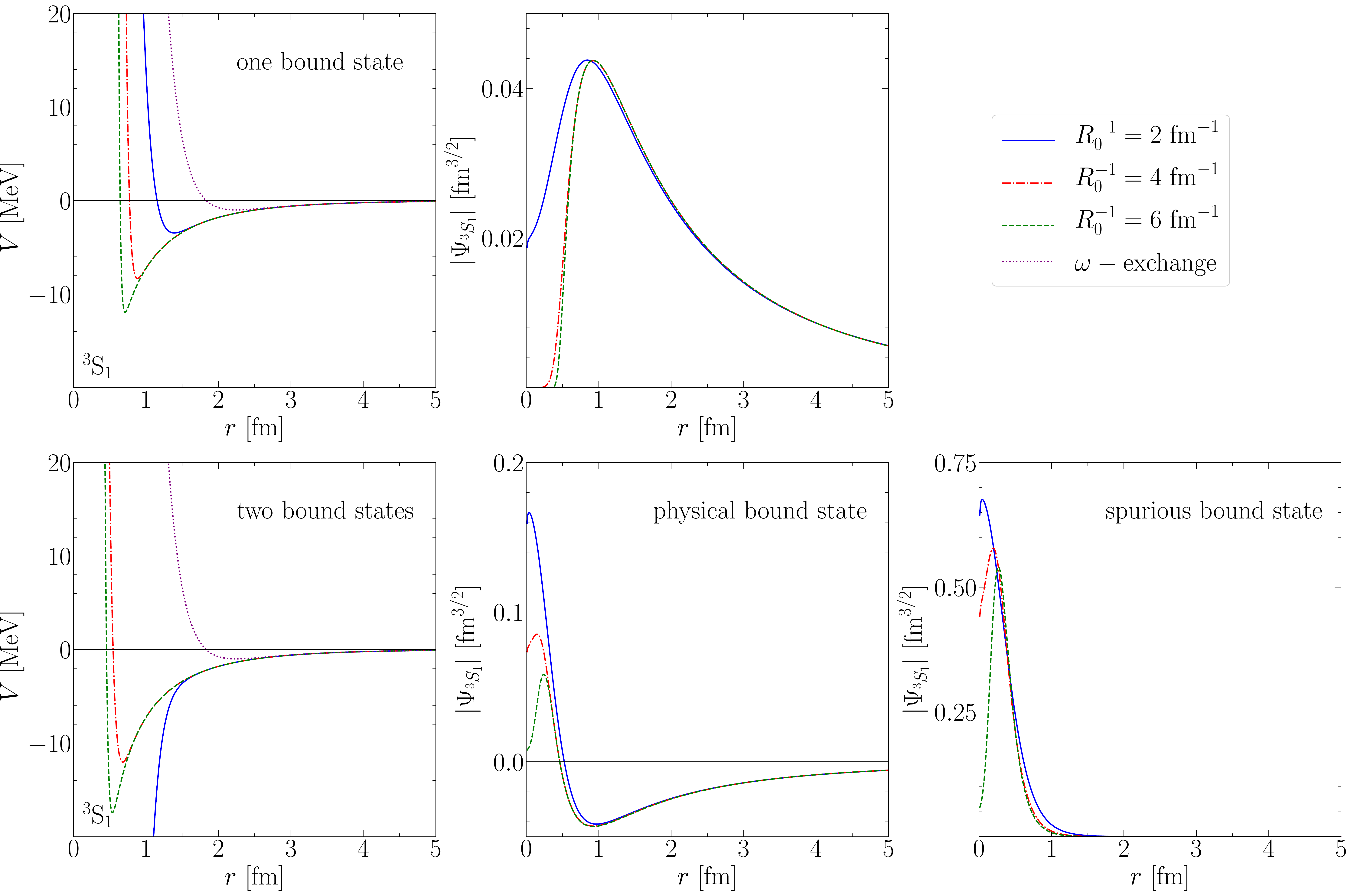}
  \caption{\label{fig:wavefun}
 Chiral LO potentials in the $^3S_1$ partial wave as well as the corresponding  
 wave functions for one or two bound states for three cutoff values and 
 $n_1=n_2=n=2$. In addition, we compare the chiral LO potentials with a potential 
 where the short-range interaction is replaced by an 
 $\omega$-meson-exchange potential.
}
\end{figure*} 

In the second panel of Fig.~\ref{fig:para} we show the LEC $C_{01}$. 
We again observe pairwise 
similar LECs for the sets $\lbrace 2,2,2\rbrace$ (blue line) and 
$\lbrace 4,1,2\rbrace$ (red line) and for the sets $\lbrace 2,2,4\rbrace$ 
(green line) and $\lbrace 4,1,4\rbrace$ (purple line), but this time we 
observe no crossing with $R_0^{-1}$. The LEC $C_{01}$ is mostly affected 
by the choice of the short-range regulator because the singular tensor 
force does not contribute to the $^1S_0$ channel and the OPE is relatively weak.
This is reflected in the LECs, which are attractive, but approach zero for 
increasing $R_0^{-1}$. The LECs can be described with high precision 
by a function of the form 
\begin{align}
C_{01}(R_0)= a R_0^b\,.
\end{align}
For the sets $\lbrace 2,2,2\rbrace$ and $\lbrace 4,1,2\rbrace$ with $n=2$, 
we find $a=-2.47 \fm^{2+b}$ and $b = -0.98$. For the sets $\lbrace 
2,2,4\rbrace$ and $\lbrace 4,1,4\rbrace$ with $n=4$, we find $a = -1.83 
\fm^{2+b}$ and $b=-0.96$. Thus, the LECs are proportional to $R_0^{-1}$.

In the third panel, we show the LEC $C_{10}$ when enforcing different 
numbers (one, two, or three) of bound states in the deuteron channel. 
As before, 
we observe a systematic behavior of the LECs with increasing $R_0^{-1}$ 
but the LECs increase much slower for interactions with more bound states. 
We note that if we allow more bound states in the deuteron channel, this 
also leads to spurious bound states in higher partial waves with spin $S = 1$.

In Fig.~\ref{fig:wavefun}, we show the potentials in the $^3S_1$ partial wave,
as well as the corresponding wave functions, for one allowed bound state 
(upper panels) and when enforcing two bound states (lower panels) for three 
cutoff values. In the former case, for increasing $R_0^{-1}$, the OPE 
extends to smaller distances, but is cut off at small $r$ by the repulsive short-range
contact interaction. When increasing $R_0^{-1}$ the range of the short-range 
regulator function decreases as expected, but this is compensated by very large LECs. This results in a hard core which does not vanish even for 
large values of $R_0^{-1}$, but cuts off the long-distance singular OPE in such 
a way that only one bound state can be accommodated. 

\begin{figure*}[t]
\includegraphics[width=\textwidth]{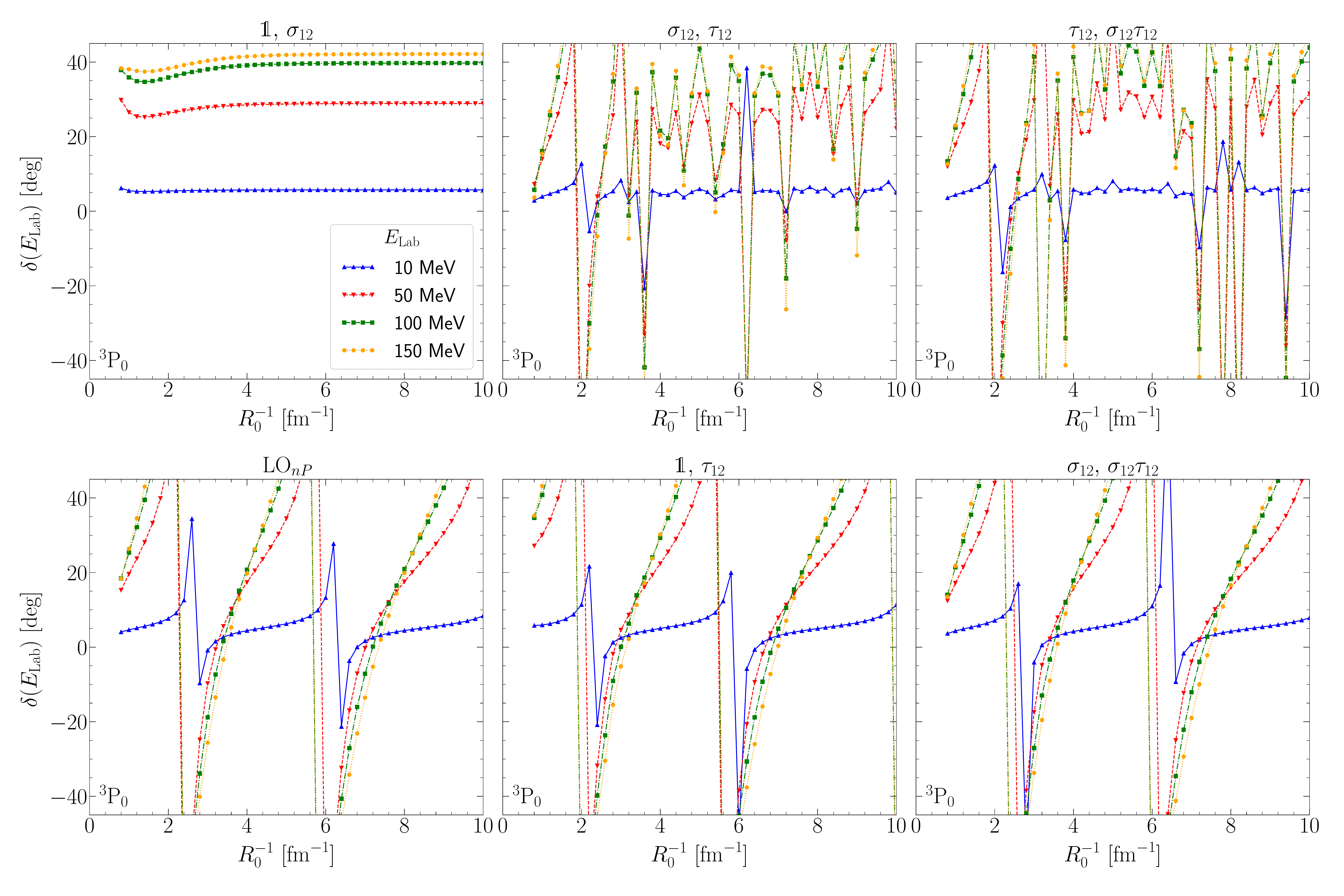}
  \caption{\label{fig:ps_fierz}
Phase shifts in the ${}^3P_0$ partial wave at laboratory energies 
$E_{\text{lab}}=$ 10, 50, 100, and 150 MeV vs 
the inverse cutoff $R_0^{-1}$ for different 
LO operator structures and $n_1=n_2=n=2$. The first panel shows the 
phase shifts for the 
operator pair $\mathds{1}$, $\sigma_{12}$, where the spin-isospin LEC 
$C_{11}$ is equal to $C_{10}$. The second and third panels show the 
operator combinations $\sigma_{12}$, $\tau_{12}$ and $\tau_{12}$, 
$\sigma_{12}\tau_{12}$, where $C_{11} \sim -C_{10}$ 
(see Table~\ref{tab:LECproj}). In the lower panels, we show the interaction 
LO$_{{\rm n}P}$, where $C_{11} = 0$ and which is closest to an interaction 
that respects Fierz rearrangement freedom, as well as the operator 
combinations $\mathds{1}$, $\tau_{12}$ and $\sigma_{12}$, 
$\sigma_{12}\tau_{12}$, where $C_{11} \sim C_{01}$ with 
$C_{01} \to 0$ (see Fig.~\ref{fig:para}). In the lower panels, we find a 
behavior similar to that in Ref.~\cite{Nogga:2005hy}.}
\end{figure*} 

These considerations allow us to understand the large-cutoff behavior of 
$C_{10}$ in Fig.~\ref{fig:para}. To cut off the OPE at a certain radius scale 
$r^{*,i}$ so that only $i$ bound states appear requires the short-range part
of the potential at this scale to be sufficiently repulsive. For $n=2$, we have
\begin{equation}
V_{10}(r^{*,i})= C_{10}\frac{1}{\pi^{\frac32}R_0^3}\exp\left(-\left(r^{*,i} R_0^{-1}\right)^2 \right)=V_0^i\,,
\end{equation}
where $V_{10}$ is the potential in the $^3S_1$ channel and $V_{0}^i$ is 
the strength necessary to compensate the OPE at $r^{*,i}$. Then, it is easy 
to see that 
\begin{equation}
C_{10}=  V_0^i \pi^{\frac32}R_0^3\exp\left(\left(r^{*,i} R_0^{-1}\right)^2 \right)\,,
\end{equation}
so $C_{10}$ has to grow exponentially in $R_0^{-2}$, which is what we 
observe in Fig.~\ref{fig:para}. In fact, fitting the large-cutoff behavior for 
$C_{10}$ in the case of one bound state, we find $C_{10}\sim \exp
\left((r^{*,1} R_0^{-1})^b \right)$ with $b\approx 1.97$ and 
$r^{*,1}\approx 0.48 \fm$, where the value of $r^{*,1}$ sets the scale 
for the crossing region in 
Fig.~\ref{fig:para}. In case of two bound states, we find 
$b\approx 1.97$ as before and $r^{*,2}\approx 0.20 \fm$. 
The exponents are in very good agreement with our expectations, 
while $r^{*,1}$ and  $r^{*,2}$ are in qualitatively good agreement with 
our findings in Fig.~\ref{fig:wavefun}, especially considering the short-distance
scale in the deuteron wave function.  The resulting hard core pushes the 
deuteron 
wave function out from the center, which we see in the second panel of
Fig.~\ref{fig:wavefun}. When enforcing two bound states, instead, the OPE 
is probed also at smaller $r$, so that two bound states can be 
accommodated. In this case, $r^{*,2}$ is smaller than $r^{*,1}$, which 
leads to smaller values of the LECs (a smaller hard core) at a certain 
$R_0^{-1}$. Our results for large $R_0^{-1}$ and $n$ deuteron bound 
states show that there exists an effective cutoff $r^{*,n}$ in coordinate 
space. It would further be interesting to investigate how this behavior 
translates to nonlocal interactions. 

We emphasize that all LECs for all numbers of bound states are 
fit to reproduce $NN$ phase shifts and that this represents an ambiguity 
when fitting nuclear forces to the phase shifts. 
Because experimentally there exists only one bound state in the deuteron 
channel, in the following we require our interactions to be on the 
one-bound-state branch. 

\section{Results for phase shifts}\label{sec:phaseshifts}

\begin{figure*}[t]
\includegraphics[width=\textwidth]{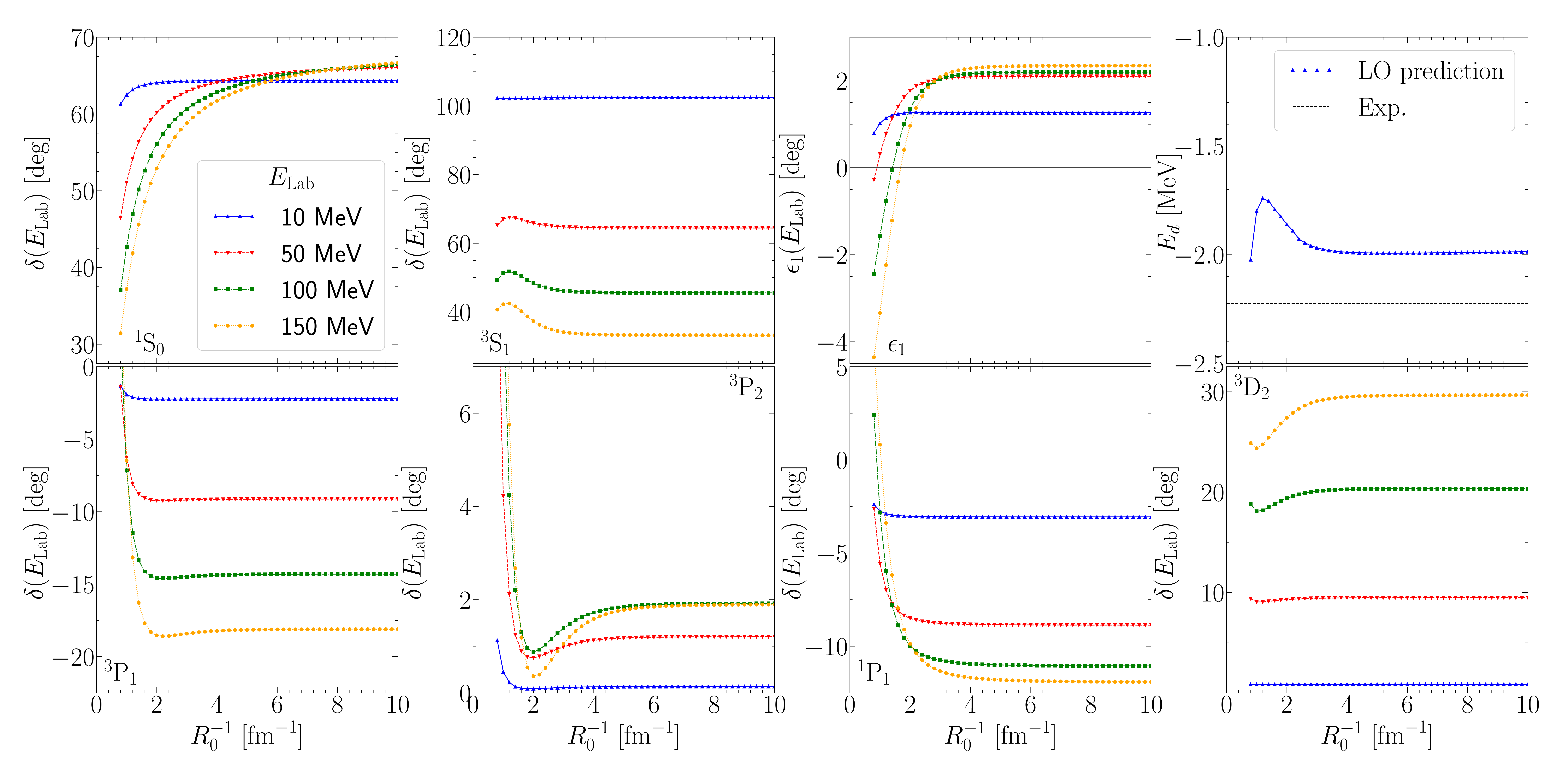}
  \caption{\label{fig:PSplot}
Phase shifts in the ${}^1S_0$, ${}^3S_1$, ${}\epsilon_1$, ${}^3D_2$, ${}^1P_1$,
${}^3P_1$, and ${}^3P_2$ partial waves for laboratory energies $E_{\text{lab}}=$ 10, 50, 100, and 150 MeV as well as the 
deuteron ground-state energy $E_d$ (upper right panel) as functions of 
the inverse cutoff $R_0^{-1}$ for the operators 
($\mathbbm{1}$, $\sigma_{12}$) and $n_1=n_2=n=2$. 
}
\end{figure*} 

Next, we investigate the phase-shift behavior as function of the
cutoff scale. We focus on the regulator with $n_1=n_2=n=2$, because this allows us to investigate the interaction 
also at large values for the inverse cutoff. For this set, we construct potentials 
for all five linearly independent operator pairs from Eq.~\eqref{eq:Loc_Cont} 
and the LO$_{\text{n}P}$ potential. Because all interactions are fit to the 
two $S$-wave channels, we obtain the same LECs $C_{10}$ and 
$C_{01}$, but the LECs $C_{00}$ and $C_{11}$ depend on the operator 
choice; see Eq.~\eqref{eq:LECtrafo} and Ref.~\cite{Huth:2017wzw}. The 
values of $C_{00}$ and $C_{11}$ for all operator pairs are listed in 
Table~\ref{tab:LECproj}, where we show both the functional dependence on 
$C_{10}$ and $C_{01}$ for each LEC, as well as the limit for $R_0\to0$ 
(see Fig.~\ref{fig:para}). The LECs $C_{00}$ and $C_{11}$ of the LO
$_{\mathrm{n}P}$ interaction are exactly zero by construction. 

In Fig.~\ref{fig:ps_fierz}, we show the $^3P_0$ phase shifts at laboratory
energies $E_{\text{lab}}=$ 10, 50, 100, and 150~MeV as function of the inverse cutoff 
for each of the operator pairs and for the LO$_{\mathrm{n}P}$ interaction. 
In general, when increasing the momentum-space cutoff, i.e., taking the 
coordinate-space cutoff $R_0\to 0$, the short-range regulator $f_s(r,R_0)$
becomes narrower. The long-range regulator $f_l(r,R_0)$, on the other hand, 
which is used in the OPE to suppress the singularity at $r=0$ while 
preserving long-range physics, allows more contributions at small distances. 
In partial waves, where the OPE tensor part is attractive and no counterterms are present, e.g., $^3P_0$, spurious bound states appear. The signature
of this effect is a limit-cycle-like behavior of the phase 
shift~\cite{Nogga:2005hy}.

\begin{table}[t]
\caption{Leading-order spin-isospin LECs $C_{ST}$ in the $S=1$, $T=1$ and 
$S=0$, $T=0$ channels as functions of the coupling constants $C_{10}$ and 
$C_{01}$ for different operator choices. For LO$_{{\rm n}P}$, 
$C_{11} = C_{00} = 0 $ by definition. The second columns indicate the limit 
for $R_0\to 0$.}\label{tab:LECproj}
\begin{tabular}{c|c|c|c|c}
\hline 
\hline 
\multirow{2}{*}{Operators} & \multicolumn{2}{c|}{$C_{11}$} & \multicolumn{2}{c}{$C_{00}$} \\ 
 &	\multicolumn{1}{c}{Exact} & \multicolumn{1}{c|}{$R_0 \to 0$} &	\multicolumn{1}{c}{Exact} & \multicolumn{1}{c}{$R_0 \to 0$} \\
\hline 
$\mathds{1},\sigma_{12}$            & $C_{10}$                     & $+\infty$ & $C_{01}$ &$ 0^-$ \\ 
$\mathds{1}, \tau_{12}$             & $C_{01}$                     & $ 0^-$ & $C_{10}$& $+\infty$ \\ 
$\sigma_{12}, \tau_{12}$            & $-\frac{1}{2}(C_{10}+C_{01})$&$-\infty$ & 
$\frac{3}{2}(C_{10} + C_{01})$      &$+\infty$ \\ 
$\sigma_{12}, \sigma_{12}\tau_{12}$ & $-\frac{1}{3}C_{01}$         &$ 0^+$ & $-3C_{10} $&$-\infty$ \\ 
$\tau_{12}, \sigma_{12}\tau_{12}$   & $-\frac{1}{3} C_{10} $       & $-\infty$& $-3C_{01} $&$ 0^+$ \\ 
\hline 
\hline 
\end{tabular} 
\end{table}

In the first row, we show the operator pairs $(\mathds{1},\sigma_{12})$,  
$(\sigma_{12}, \tau_{12})$, and $(\tau_{12},\sigma_{12}\tau_{12})$, for 
which $C_{11} \sim C_{10}$. For the pair $(\mathds{1},\sigma_{12})$, 
the LEC is repulsive $C_{11}> 0$ and thus acts to compensate the attractive 
tensor contribution from the OPE interaction. In this case, we find the results 
stabilize on plateaus when the cutoff is increased. This is the only operator 
pair for which we observe that the phase shifts become independent of 
$R_0^{-1}$ for large $R_0^{-1}$. For the other two operator pairs, the 
corresponding LEC is attractive, $C_{11} < 0$, which adds to the OPE attraction and leads to the appearance of spurious bound 
states in the $^3P_0$ wave. This causes a highly oscillatory behavior of the 
phase shifts.

\begin{figure*}[t]
\includegraphics[width=\textwidth]{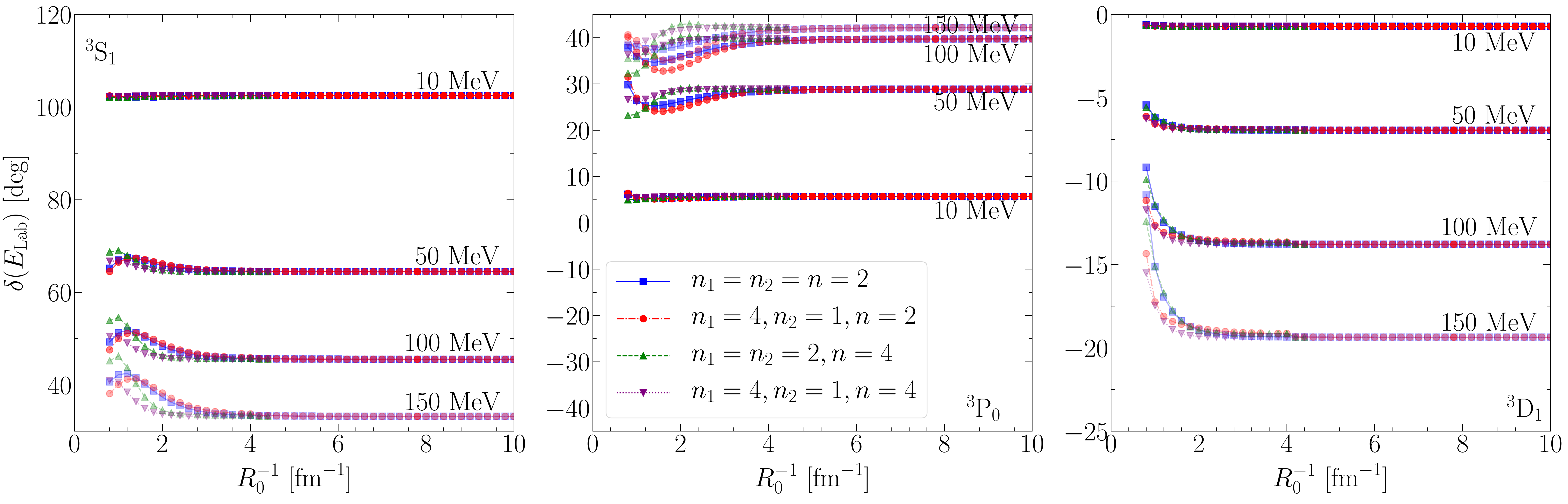}
  \caption{\label{fig:diffreg}
 Phase shifts in the $^3S_1$, $^3P_0$, and $^3D_1$ partial waves for
 different laboratory energies and for different regulator functions, as 
 functions of the inverse cutoff $R_0^{-1}$ for the operators 
 $(\mathds{1}, \sigma_{12})$.
}
\end{figure*} 

In the second row, we show the LO$_{\mathrm{n}P}$, $(\mathds{1},
\tau_{12})$, and $(\sigma_{12},\sigma_{12}\tau_{12})$ interactions 
for which $C_{11} \to 0$. For the LO$_{\mathrm{n}P}$ interaction, the $P$
waves only receive contributions from OPE, and thus 
this interaction has the closest resemblance to nonlocal chiral EFT 
interactions (i.e., to the case studied in Ref.~\cite{Nogga:2005hy}). The other 
two interactions
lead to short-range contributions in the $^3P_0$ wave, but these are small, 
and the overall phase shifts are very similar to the LO$_{\mathrm{n}P}$ 
interaction. Phase-shift jumps in Fig.~\ref{fig:ps_fierz} correspond to cutoff 
values where new bound states enter in the $^3P_0$ wave. For the 
interactions in the lower panels, we find a limit-cycle-like behavior similar to the 
nonlocal potentials of Ref.~\cite{Nogga:2005hy} without counterterms.

For the interaction with operators  $(\mathds{1},\sigma_{12})$, which 
leads to plateaus, we show the phase shifts in the $^1S_0$, $^3S_1$, 
$^3P_1$, $^3P_2$, $^1P_1$, and $^3D_2$ partial waves as well as the 
mixing angle $\epsilon_1$ and the deuteron ground-state energy in 
Fig.~\ref{fig:PSplot}, and find plateaus in all cases for 
$R_0^{-1}\gtrsim 4 \fm^{-1}$, similar to Ref.~\cite{Nogga:2005hy} when
counterterms were included there. At 
higher laboratory energies, the plateau is reached for higher values of 
$R_0^{-1}$. The plateau values for the phase shifts are very similar to the 
results found by NTvK~\cite{Nogga:2005hy}, except in the attractive-tensor 
$P$ and $D$ waves, without adding any new counterterms, in contrast 
to NTvK. For the deuteron, we find a ground-state energy of 
$E_d\to -1.99 \mev$ for $R_0^{-1}\to \infty$, which is close to experiment
although the deuteron was not included in the fit.

\begin{figure*}[t]
\includegraphics[width=\textwidth]{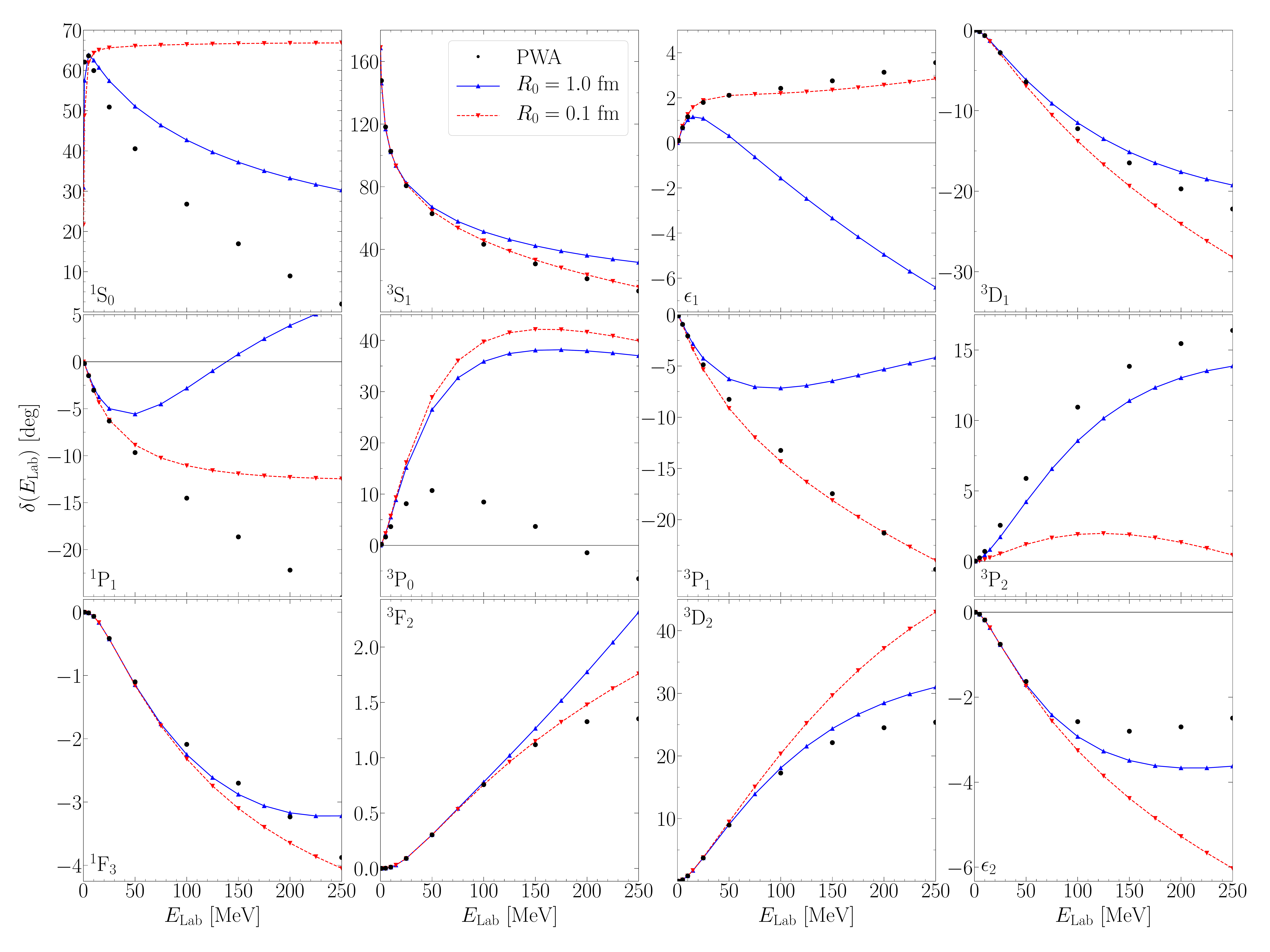}
  \caption{\label{fig:ps}
Phase shifts in the ${}^1S_0$, ${}^3S_1$, ${}\epsilon_1$, ${}^3D_1$, 
${}^1P_1$, ${}^3P_{0,1,2}$, ${}^1F_3$, ${}^3F_2$, $\epsilon_2$, and 
${}^3D_2$ channels as function of the laboratory energy for the inverse cutoffs $R_0^{-1}=1\!\fm^{-1}$ and 
$R_0^{-1} = 10\!\fm^{-1}$ for $n_1=n_2=n=2$ in comparison to the PWA 
phase shifts.
}
\end{figure*} 

In Fig.~\ref{fig:diffreg}, we show the phase shifts as function of $R_0^{-1}$ 
in three partial waves for different laboratory energies and different regulator 
choices. We find that the phase shifts converge to the same values, and that 
the plateaus are independent of the exponents in the regulator functions. Note 
that for the sets $\lbrace n_1,n_2,n\rbrace=\lbrace 2,2,4\rbrace$ and  $
\lbrace 4,1,4\rbrace$ we do not obtain numerical results for 
$R_0^{-1} > 4.4\ \mathrm{fm}^{-1}$ as discussed for Fig.~\ref{fig:para}.

The phase shift plateaus do not necessarily have to be close to the physical phase shift values. In Fig.~\ref{fig:ps}, we compare the phase shifts 
as function of the laboratory energy in several partial waves for two cutoff scales ($R_0^{-1} =1.0$ fm$^{-1}$ and 
$R_0^{-1} =10$ fm$^{-1}$) with the PWA 
values. The large-cutoff interaction that lies on the 
phase-shift plateaus describes the energy dependence of the phase shifts 
reasonably well and in some cases much better than the result for a typical 
low cutoff. The only exception is the $^1S_0$ partial wave, because at
LO the effective range cannot be correctly described. It is not clear if an
improvement can be found at NLO due to causality 
bounds~\cite{Phillips:1996ae}, and it will be interesting to investigate the 
order-by-order behavior at large $R_0^{-1}$. 

\section{Discussion}\label{sec:discussion}

Because of the attractive singular OPE, results are very cutoff dependent
in WPC without the promotion of additional counterterms. In this paper, we 
have explained the fact 
that local regulators connect the LO counterterms with all higher 
partial waves. For a certain class
of local regulators, these regulator artifacts can compensate the attractive 
tensor contributions from OPE and we find cutoff-independent results. 
However, we state explicitly that these results do not imply that WPC is renormalizable.

We investigated an interesting case for local interactions, which may be 
beneficial from a practical point of view as local interactions with larger 
$R_0^{-1}$ can easily be explored with quantum Monte Carlo methods and 
may allow to reduce regulator artifacts in many-body systems. In 
Ref.~\cite{Tews:2015ufa}, it was found that lowering the $3N$ cutoff 
$R_{3N}$ in pure neutron matter while keeping a constant $NN$ cutoff 
$R_0$  leads to collapses of the many-body system. The best result was 
found when $R_{3N}=R_0$. If it is possible to construct chiral interactions 
with smaller $NN$ cutoff and no spurious bound states, as we have shown 
here, one can lower the $3N$ cutoff while at the same time avoiding such 
collapses. This will help to reduce $3N$ regulator artifacts, which have 
been found to be sizable; see Refs.~\cite{Dyhdalo:2016ygz, 
Tews:2018kmu}. This may therefore allow to significantly reduce uncertainties 
in many-body calculations with local chiral interactions. The results of such 
calculations will be reported in a future paper. 

Conceptually, our results are very different from the results of NTvK. While 
NTvK restore renormalizability of chiral interactions at LO by adding additional 
counterterms in channels with attractive tensor interactions and then obtain 
cutoff-independent results, we have seen that the regulator artifacts for 
local chiral interactions with the operator combination 
($\mathbbm{1},\ \sigma_{12}$) add additional repulsion to the same partial waves, which in 
turn leads to similar plateaus; see Fig.~\ref{fig:ps_fierz}. However, the 
appearance of plateaus does not mean that our interactions are 
renormalizable.

Our findings are only possible for certain local regulators that are smooth
functions. Furthermore, the appearance of plateaus is only possible 
because we require the interactions to remain on the branch for a single bound state in the deuteron channel. . As we have explored before, in this case the width 
of the short-range regulator function decreases with $R_0^{-1}$ but this is 
compensated by increasing values of the LECs. This results in a nonvanishing 
hard core which compensates the attraction from OPE in such a way that only 
one bound state can be accommodated. 

The core is strong enough to counter the attraction in the
$^3S_1-^3D_1$ channel (where it is strongest), and it is only natural that 
it is sufficiently strong to counter the attraction in higher partial waves, 
where we find no additional spurious bound states. This is another difference 
from the results of NTvK, where the introduction of additional counterterms 
does not eliminate the appearance of spurious bound states. If we instead 
allow for more bound states to appear in the $^3S_1$ channel, the values of 
the LECs decrease and the core is reduced in magnitude. Then, we also
find a limit-cycle cutoff dependence behavior and additional counterterms 
need to be added in the attractive tensor channels, as NTvK found. 

We remark, however,  that if we would allow more bound states 
to enter, this would mean that the LECs $C_{ST}(R_0)$ have to jump from 
one bound-state branch to another bound-state branch. To our knowledge, in 
this case there is no clear prescription at which cutoff values these jumps 
have to occur, which introduces an additional ambiguity. Enforcing one bound 
state, on the other hand, is a reasonable and practical prescription for the 
construction of potentials that introduces no new ambiguity.

While the connection of the $S$-wave contact interactions with higher 
partial waves is purely a regulator artifact, such a connection can have 
a qualitative physical motivation. The LO counterterms in WPC absorb, 
among others, the effects of heavier mesons like the $\rho$ or $\omega$ 
meson, which are integrated out in chiral EFT~\cite{Epelbaum:2001fm}. 
Such heavier mesons are responsible for the 
short-range $NN$ repulsion and compensate the singular attraction of the 
OPE interaction; see also Ref.~\cite{Beane:2008bt}. The exchange 
interactions for 
these mesons are local in the static limit and enter all partial waves, similarly 
to pion exchanges. Local regulators establish a similar connection of all 
partial waves. 

For example, the central short-range repulsion in one-boson-exchange
potentials originates from $\omega$-meson 
exchange~\cite{Machleidt:1987hj}. The exchange potential of a very heavy 
$\omega$-like vector meson would be given by
\begin{align}
V_{\omega}(m_{\omega},r)&\stackrel{m_{\omega}\gg m_N}{\longrightarrow} \frac{e^{-M_{\omega} r}}{r} \left \lbrace
\left(\mathbbm{1}+\frac13 \sigma_{12}  \right) \right.\\
&\quad \left. -\frac16 \left(1+\frac{3}{M_{\omega} r}+\frac{3}{(M_{\omega} r)^2} \right)  S_{12}({\bf r})\right\rbrace \nonumber\,
\end{align}
with the mass of the meson $m_{\omega}$ and the nucleon mass $m_N$; see, e.g., Eq.~(F.8) in Ref.~\cite{Machleidt:1987hj}. In the case of the local 
LO potential with the operator structure ($\mathbbm{1}, \sigma_{12}$), 
we find that the operator LECs for larger cutoffs approach $C_\mathds{1}= 
3C_\sigma$, because $C_{01} = C_\mathds{1} - 3 C_\sigma\to 0$. 
Thus, a corresponding meson would need to have the leading operator 
structure $\sim \mathbbm{1}+1/3 \sigma_{12}$, which is exactly the 
central part of an $\omega$-like vector meson. We highlight this in the 
left panels of Fig.~\ref{fig:wavefun}, where, in addition to the results for the central chiral interactions, we also show the result for the $\omega$-meson exchange. 
We find a behavior similar to the chiral interactions for larger cutoffs. 

\section{Summary}\label{sec:summary}

In this paper, we have investigated the behavior of local chiral
interactions in WPC when the coordinate-space cutoff is lowered. We have 
constructed LO interactions for cutoffs ranging from $R_0^{-1} = 0.8$ to $10.0\ \mathrm{fm}^{-1}$ for different choices of 
the regulator function and for different pairs of the LO operators of 
$\left\lbrace \mathds{1}, \sigma_{12}, \tau_{12}, \sigma_{12}\tau_{12}\right\rbrace$. Our interactions were fit to reproduce the $S$-wave phase 
shifts from the Nijmegen PWA. We additionally required only one bound state 
in the deuteron channel but did not use the deuteron binding energy to 
constrain our fits.

Our results show that, for the operator combination $(\mathds{1}, \sigma_{12})$, phase shifts in all partial waves as well as the deuteron 
ground-state energy exhibit a plateau when increasing the inverse cutoff
$R_0^{-1}$, leading to cutoff-independent results. This can be explained because in the fit 
the attractive tensor contribution from OPE in the deuteron channel is 
compensated by the short-range contact interactions to guarantee only one 
bound state. Local regulators mix contributions from all operators into all 
partial waves, i.e., LO operators that nominally only describe $S$-wave physics 
contribute to all channels once a local regulator is used. For the operator choice 
$(\mathds{1}, \sigma_{12})$, the LEC $C_{10}$ in the $^3S_1$ channel is 
mixed into all attractive tensor channels with the same sign, providing sufficient 
repulsion in these channels. 

Thus, using the artifacts of local regulators to our advantage allowed us 
to construct interactions that enable a cutoff-independent description of 
phase shifts.
Comparing the phase-shift predictions for these hard interactions 
with phase shifts from the PWA, we found very good agreement at LO.

We state again that these results do not imply that WPC is renormalizable. 
However, our findings may prove useful from a practical point of view, as they
may allow to reduce regulator artifacts in many-body calculations. In the 
future, we will investigate these hard potentials in quantum Monte Carlo
calculations of nuclear systems to investigate if this behavior persists.

\begin{acknowledgments}
We thank R. Furnstahl, A. Gezerlis, D. B. Kaplan, S. K\"onig, J. Lynn, 
D. Philips, U. van Kolck, and C. J. Yang for useful discussions.
This work was supported in part by the National Science Foundation  
Grant No. PHY-1430152 (JINA Center for the Evolution of the Elements), 
the European Research Council Grant No. 307986 STRONGINT, the Deutsche 
Forschungsgemeinschaft Grant No. SFB 1245, and the U.S.~Department of 
Energy Grant No.~DE-FG02-00ER41132.
Computational resources have been provided by the J\"ulich
Supercomputing Center and the Lichtenberg high-performance computer of TU Darmstadt.
\end{acknowledgments}

\end{document}